\documentclass[prc,twocolumn,showpacs,amssymb]{revtex4}
\usepackage{graphicx}

\begin{document}

\title{Evolution of single-particle  states beyond $^{132}$Sn }

\author{L. Coraggio$^{1}$, A. Covello$^{1,2}$, A. Gargano$^{1}$,
and N. Itaco$^{1,2}$} 
\affiliation{$^{1}$Istituto Nazionale di Fisica Nucleare, 
Complesso Universitario di Monte S. Angelo, I-80126 Napoli,
Italy\\
$^{2}$Dipartimento di Fisica, Universit\`a
di Napoli Federico II,
Complesso Universitario di Monte S. Angelo,  I-80126 Napoli,
Italy}

\date{\today}

\begin{abstract}

We have performed shell-model calculations for the two one valence-neutron isotones $^{135}$Te and $^{137}$Xe and the two one valence-proton 
isotopes $^{135,137}$Sb. The
main aim of our study has been to investigate the evolution of single-particle 
states with increasing nucleon number. To this end, we have  focused attention 
on the spectroscopic factors and the effective single-particle energies.
In our calculations, we have employed a realistic low-momentum two-body
effective interaction derived from the CD-Bonn nucleon-nucleon
potential that has already proved quite successful in describing the 
spectroscopic properties of nuclei in the $^{132}$Sn region. Comparison shows 
that our results reproduce very well the available experimental data. 
This gives confidence in the evolution of the single-particle states 
predicted by the present study.

\end{abstract}    

\maketitle

\section{Introduction}

The evolution of single-particle states when moving away from double shell closures has long been a subject of primary interest in nuclear structure physics. It was in the early 1960s \cite{Cohen61}, in fact, that quantitative experimental information on this subject started to become available thanks to the study of one-nucleon transfer reactions such 
as ($d$,$p$) and ($d$,$t$). During the 1960s and early 1970s  the field developed rapidly and spectroscopic factors were extracted from pickup and stripping reactions for many
nuclei. In particular, the stable Sn isotopes and $N=82$ isotones were the subject of extensive studies \cite{Schneid67,Wildenthal71} which provided relevant information on the shell-model structure of these nuclei. By the end of the 1970s, however, most of the feasible experiments
had been performed and the study of transfer reactions began loosing its role as a most powerful tool for the understanding of nuclear structure. 

As it was to be expected, the advent of radioactive ion beams (RIBs) has opened a new era for the study of transfer reactions, which is currently recognized as one of the major research themes to be pursued at the second-generation RIB facilities.

A main feature of transfer reactions with RIBs is that they have to be performed in inverse kinematics. This is for instance
the case of the $^{132}$Sn($d$,$p$) reaction recently performed at the first-generation RIB facility at Oak Ridge National Laboratory \cite {Jones10,Jones11}. This experiment allowed to
investigate the single-particle structure of the one-valence neutron nucleus $^{133}$Sn, and the spectroscopic factors for the ground and three excited states extracted from the differential cross sections turned out to be very close to one. This is a remarkable achievement which has set the stage for future studies of neutron-rich nuclei in the $^{132}$Sn region, which are also of great interest for modelling the $r$ processes of nucleosynthesis.

In this context, it is worth mentioning that it is also of current interest the
 study of transfer reactions with stable beams in inverse kinematics to test new detectors specifically designed  
for use with heavy beams. A recent notable example of this kind of study is the 
$^{136}$Xe ($d$,$p$) reaction  performed at the ATLAS facility at Argonne National Laboratory
\cite{Kay11}, which has provided a stringent test of the HELIOS spectrometer used to analyze the outgoing protons. 

The experimental studies of nucleon transfer reactions currently going on and the bright perspectives for future research with RIBs are of course a great stimulus for theoretical
studies aimed at understanding the evolution of single-particle states. This is indeed the case of our very recent shell-model study \cite{Coraggio13} of the single-neutron properties of
$^{137}$Xe, which received motivation from the experiment of Ref. \cite{Kay11}.

Based on the results obtained in \cite{Coraggio13}, we have found it very interesting to perform a similar study for other nuclei beyond  $^{132}$Sn which are 
within reach of ($d$,$p$) transfer reactions with RIBs. More precisely, we consider
the two one valence-neutron isotones $^{135}$Te and $^{137}$Xe and  the
two one valence-proton isotopes $^{135,137}$Sb. Actually, results for 
$^{137}$Xe  have 
already been given in \cite{Coraggio13}, where we essentially focused on the 
comparison between our spectroscopic factors and those reported 
in~\cite{Kay11}. The present paper is of more general scope, in that we study 
how the single-particle states of $^{133}$Sb and $^{133}$Sn are modified by 
the addition of pairs of neutron and protons, respectively.
 
The article is organized as follows. In Sec. II we give an outline of our calculations. In Sec. III the results for the energies and spectroscopic factors are presented and discussed in detail. In Sec. IV we present our predicted effective single-particle energies and discuss their evolution. Sec. V contains a summary of our conclusions.

\section{Outline of calculations}

The results presented in this section have been obtained with the  same  Hamiltonian used in our previous shell-model studies of neutron-rich nuclei beyond $^{132}$Sn \cite{Coraggio09,Covello09,Covello11}. In this Hamiltonian, the single-particle energies are taken from experiment while the two-body effective interaction  is derived within the framework of
perturbation theory~\cite{Coraggio09b,Coraggio12} starting  from the 
CD-Bonn $NN$ potential  \cite{Machleidt01} renormalized by way of the $V_{\rm low-k}$ 
approach~\cite{Bogner02}. 
Some details on the calculation of this  two-body interaction can be found in 
\cite{Coraggio13}. We discuss here only  the choice of the single-particle energies that may be relevant to the  discussion of our results.

\begin{table}
\caption{Single-proton and -neutron energies (in MeV).} 

\begin{ruledtabular}
\begin{tabular}{cccc}
$\pi(n,l,j)$&$\epsilon$ & $\nu (n,l,j)$& $\epsilon$\\
\colrule
$0g_{7/2}$ & 0.00 & $1f_{7/2}$ & 0.00\\
$1d_{5/2}$ & 0.962 & $2p_{3/2}$ & 0.854 \\
$2d_{3/2}$ & 2.440 & $2p_{1/2}$ & 1.363\\
$0h_{11/2}$ & 2.792  & $0h_{9/2}$ & 1.561\\
$2s_{1/2}$ & 2.800 & $1f_{5/2}$ & 2.005\\
&   &                $0i_{13/2}$ & 2.690  \\
\end{tabular}
\end{ruledtabular}
\label{tabsp}
\end{table}

We assume that the valence protons and neutrons occupy the first major shells 
outside doubly magic $^{132}$Sn. Namely, protons are allowed to move in the five 
orbits $0g_{7/2}$, $1d_{5/2}$, $1d_{3/2}$,  $0h_{11/2}$, and   
$2s_{1/2}$ of the 50-82 shell and neutrons in  the six orbits $1f_{7/2}$, 
$2p_{3/2}$, $0h_{9/2}$,  $2p_{1/2}$,
$1f_{5/2}$,  and $0i_{13/2}$ of the
82-126 shell. 

The adopted single-proton and -neutron energies are shown in Table~\ref{tabsp}.
For protons, the first four are taken to be  
the experimental energies of the levels with corresponding spin and parity in  
$^{133}$Sb~\cite{ENSDF}. The energy of the  $s_{1/2}$ orbit, which  is still 
missing,  is taken from ~\cite{Andreozzi97}, where it was determined by 
reproducing the experimental energy of the $1/2^+$ state at  2.150 MeV in 
$^{137}$Cs. This is, in fact, populated with significant strength in 
the $^{136}$Xe($^{3}$He,~$d$)$^{137}$Cs transfer reaction~\cite{Wildenthal71}, 
while no $1/2^+$  state has been yet  identified in the lighter  $N=82$ 
isotone, $^{135}$I. 
It is worth noting that no spectroscopic factors are available for the states 
of $^{133}$Sb, which has been mainly studied by $\beta$-decay experiments 
in the '90s~\cite{Sanchez99}.

As already mentioned in the Introduction, the single-neutron  nature of the observed states in $^{133}$Sn has been recently studied in a ($d$,$p$) transfer 
reaction with a $^{132}$Sn RIB \cite{Jones10,Jones11}.  In this experiment, the 
spectroscopic factors of the previously 
observed~\cite{Hoff96} $7/2^-$,  $3/2^-$, and $5/2^-$ states were extracted 
evidencing  a little fragmentation of the single-particle strength. Furthermore, a 
strong candidate for the $2p_{1/2}$ single-particle orbit was identified
at 1.363 MeV excitation energy,  which is about 300 keV lower than the previous
proposed value~\cite{Hoff96}. 
A state at 1.561 MeV  was not significantly populated in this ($d$,$p$) reaction. This state, associated with $J^{\pi}=9/2^-$ in previous $\beta$  decay \cite{Hoff96} and spontaneous fission~\cite{Urban99} experiments, is expected to correspond to  
the  $0h_{9/2}$ orbit.
The $13/2^+$ state has not yet been observed  in $^{133}$Sn and the energy reported in Table~\ref{tabsp}  has been estimated from that (2.434 MeV) of  the 
experimental  $10^+$ state in $^{134}$Sb, assumed to belong to the $\pi g_{7/2} \nu i_{13/2}$ 
configuration. 

The shell-model calculations have been performed by using the OXBASH 
code~\cite{OXBASH}.

\section{Energies and spectroscopic factors: results and discussion}

In this section,  we present and discuss our results for $^{135}$Te and 
$^{137}$Xe with two and four protons  in addition to 
$^{133}$Sn, as well as those for 
$^{135}$Sb and $^{137}$Sb with two and four neutrons in addition to $^{133}$Sb.
 We focus attention on the  energies and  spectroscopic factors of states with angular 
momentum and parity corresponding to those of the single-neutron and -proton 
orbits for  $N=83$ isotones and Sb isotopes, respectively. In fact, it is a main aim
of our study to find out whether states with single-particle
 character  survive when  adding nucleon pairs and to which extent this 
depends on the nature of the added pairs.

\subsection{$^{135}$Te  and $^{137}$Xe}
Let us start with the $N=83$ isotones. Excitation energies and spectroscopic factors for $^{135}$Te and $^{137}$Xe are shown in  Tables~\ref{135te} and ~\ref{137xe}, respectively, where only states with $C^{2}S \geq 0.07$ are 
included.
 In Table~\ref{137xe}, we also show  the  energies and spectroscopic factors 
obtained in~\cite{Kay11} for levels  which can be safely identified with the 
calculated ones.  For a detailed comparison between experiment and theory,
we refer to \cite{Coraggio13} while here  
some points  relevant to the present discussion will be considered.
For $^{135}$Te only experimental energies  are  available~\cite{ENSDF}, although  a ($d$,$p$) 
reaction on $^{134}$Te was performed at HRIBF~\cite{Cizewski09}. In fact, preliminary 
tentative results from this experiment do not allow to extract spectroscopic 
factors but only to identify the $7/2^-$ ground state and the 
 $1/2^-$ and  $3/2^-$ yrast states as resulting from the $l=3$ and 1 transfers,
respectively. 
\begin{table}
\caption{Calculated energies and spectroscopic factors for states in $^{135}$Te. The available experimental data are reported for comparison (see text for details).} 
\begin{ruledtabular}
\begin{tabular}{ccccc}
\multicolumn{3} {c} {Calc.}  &  \multicolumn{2} {c}
 {Expt.} \\
 \cline{1-3}    \cline{4-5}
$J^{\pi}$& E(MeV) & $C^{2}S$ & $J^{\pi}$ & E(MeV) \\
\colrule
($1/2^-$)$_1$ & 1.110 & 0.45 & ($1/2^{-}$) &  1.083 \\
($1/2^-$)$_2$ & 1.947 & 0.32 &  &    \\
($1/2^-$)$_3$ & 2.400 & 0.16 &  &    \\
($3/2^-$)$_1$ & 0.726 & 0.63 & ($3/2^{-}$)  &  0.659 \\
($3/2^-$)$_2$ & 1.721 & 0.27 &  &  \\
($5/2^-$)$_1$ & 1.119 & 0.12 & ($5/2^{-}$) &  1.127 \\
($5/2^-$)$_6$ & 2.238 & 0.41 &  &   \\
($7/2^-$)$_1$ & 0.000 & 0.88 & ($7/2^{-}$)  &  0.000 \\
($9/2^-$)$_1$ & 1.302 & 0.18 & ($9/2^{-}$) &  1.246 \\
($9/2^-$)$_2$ & 1.346 & 0.51 & ($7/2^{-}$,$9/2^{-}$)&  1.380 \\
($9/2^-$)$_5$ & 2.214 & 0.07 & &   \\
($9/2^-$)$_6$ & 2.308 & 0.07 & &   \\
($13/2^+$)$_1$ & 2.268 & 0.72 & ($13/2^{+}$) &  2.109 \\      
($13/2^+$)$_2$ & 3.356 & 0.19 &  &  \\      
\end{tabular}
\end{ruledtabular}
\label{135te}
\end{table}

\begin{table}
\caption{Calculated energies and spectroscopic factors for states in $^{137}$Xe. The available experimental data are reported for comparison (see text for details).} 
\begin{ruledtabular}
\begin{tabular}{cccccc}
\multicolumn{3} {c} {Calc.}  & \multicolumn{3} {c}
 {Expt.} \\
 \cline{1-3}     \cline{4-6}
$J^{\pi}$& E(MeV) & $C^{2}S$ & $J^{\pi}$ & E(MeV)  & $C^{2}S$ \\
\colrule
($1/2^-$)$_1$ & 1.127 & 0.43 & $1/2^{-}$,$3/2^{-}$ &  0.986 & 0.35\\
($1/2^-$)$_2$ & 1.926 & 0.13 &  &  &  \\
($1/2^-$)$_4$  & 2.305 & 0.18 &  &  & \\
($1/2^-$)$_5$  & 2.407 & 0.07 &  &  & \\
($3/2^-$)$_ 1$& 0.728 & 0.57 & $3/2^{-}$ &  0.601 & 0.52\\
($3/2^-$)$_ 2$& 1.708 & 0.07 & &   & 0\\
($3/2^-$)$_ 3$ & 1.783 & 0.15 &  &  & \\
($5/2^-$)$_1$ & 1.349 & 0.17 & $5/2^{-}$ &  1.303 & 0.22\\
($5/2^-$)$_5$  & 2.039 & 0.20 &  &  &  \\
($7/2^-$)$_1$ & 0.000 & 0.86 & $7/2^{-}$ &  0.000 & 0.94\\
($9/2^-$)$_1$ & 1.327 & 0.72 & $9/2^{-}$ &  1.218 & 0.43\\
($13/2^+$)$_1$ & 2.082 & 0.75 & ($13/2^{+}$) &  1.751 & 0.84\\      
\end{tabular}
\end{ruledtabular}
\label{137xe}
\end{table}

From Tables~\ref{135te} and ~\ref{137xe} we see that the agreement between  experimental and calculated energies is very good. The 
largest discrepancy in both  $^{135}$Te and $^{137}$Xe is found for the  $13/2^+$ state, which is overestimated by 150 and 330 keV, respectively. In this 
connection, it is worth mentioning that its position is directly related to the
energy of the $0i_{13/2}$ level, which, as 
mentioned in the previous section, is still missing.  

Our calculations for  $^{135}$Te 
confirm  the preliminary results of ~\cite{Cizewski09}, the   $7/2^-$,  
$1/2^-$ and $3/2^-$ yrast states  having, in fact, the largest spectroscopic 
factors.
Actually,  it turns out that this  $7/2^-$ state, as well as the yrast $13/2^+$ 
state, carries the largest fraction of the single-particle strength, while
a non-negligible percentage  of the $p_{1/2}$ and $p_{3/2}$ strengths  is 
 distributed over higher-energy  states.
 As for the  $J^{\pi}=9/2^-$ states, we find that   it is the yrare state 
at 1.346 MeV to 
have the largest spectroscopic factor, 0.51,  to be compared to 0.18 for
the yrast one. These predicted two $9/2^-$ states are quite close in energy, the difference being 
only  a few tens of keV. Note that a 
state at 1.380~MeV has been observed with no firm spin assignment
($7/2^-$, $9/2^-$), which could be associated either with our second 
$9/2^-$ state or our second $7/2^-$ state at 1.336  MeV  with $C^{2}S=0.06$. 
Finally, we see that the largest strength of the $f_{5/2}$ orbit  is predicted 
to be carried by the 6th $5/2^-$ state at 2.238 MeV, $C^{2}S=0.41$, while 
the spectroscopic factors of the lowest five states  are much smaller.

When one adds two more protons going from  $^{135}$Te  to $^{137}$Xe, a state
of a single-particle nature can be still identified  for each $J^{\pi}$, which is
the yrast state, except for $J^{\pi}=5/2^-$.
As for the  $f_{5/2}$ strength, a stronger  fragmentation is found in  $^{137}$Xe 
with respect to $^{135}$Te, the highest  spectroscopic factor 
being 0.20   for the 5th excited state. Our findings for $^{137}$Xe 
are consistent with the experimental results, although the 
spectroscopic factor of the yrast $9/2^-$ state is overestimated by the 
theory. This point is discussed in Ref.~\cite{Coraggio13}.

\begin{figure}

\centering

\includegraphics [scale=0.8]{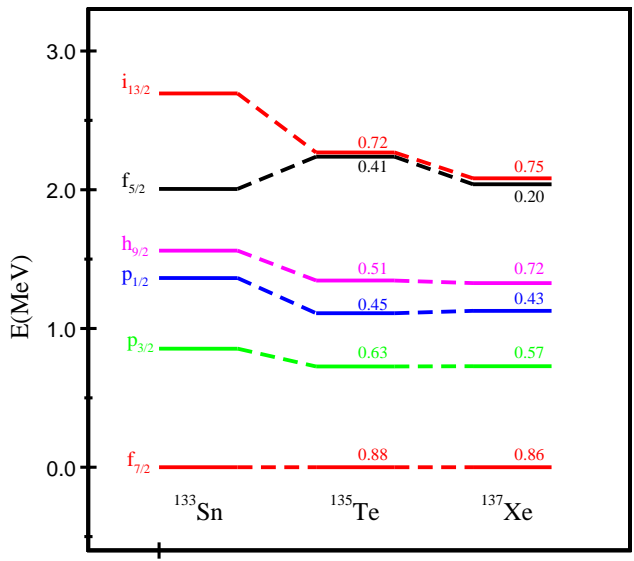}

\caption{Calculated energy levels with corresponding spectroscopic factors  in 
the $N=83$ isotones $^{135}$Te and  $^{137}$Xe (see text for details). Single-neutron levels for $^{133}$Sn are also reported.}\label{fig:1}

\end{figure}

Based on the above considerations, we have found it interesting to compare the
adopted single-neutron energies for  $^{133}$Sn  with the calculated spectra 
 of $^{135}$Te and $^{137}$Xe, including,  for each $J^{\pi}$,
only the level with the largest spectroscopic factor. 
This is done in 
Fig. \ref{fig1},  where we see that  when adding proton pairs to $^{133}$Sn the 
single-neutron spectrum preserves its original  structure, becoming  on the whole slightly more  
compressed. Actually, only the  $5/2^-$ state  in both nuclei moves up in 
energy. In this context, it may be worth mentioning that the lower $5/2^-$ 
states, 
which are not of a single-particle nature,  are predicted by our calculations 
to arise from the  
coupling of the single-neutron $f_{7/2}$ state  to two-proton excitations. 
The $2^+$, $4^+$, and $6^+$ in $^{134}$Te lie, in fact, at an energy ranging 
from 1.3 to 1.7 MeV, which is lower than that, 2.0 MeV,  of the $f_{5/2}$ level in $^{133}$Sn.

\subsection{$^{135}$Sb and $^{137}$Sb}

The energies and spectroscopic factors for $^{135}$Sb and $^{137}$Sb are 
reported
in Tables \ref{135sb} and \ref{137sb}, respectively, including
only states with $C^{2}S \geq 0.07$  selected among  the lowest twenty states for each angular momentum.  For $^{137}$Sb only the ground state is known,  while for $^{135}$Sb several states have been observed,  most of them, however,  without firm spin assignment. Furthermore, no spectroscopic factors have been obtained.
In Table~\ref{135sb}  the experimental energies of  $^{135}$Sb \cite{ENSDF} for states which can be safely associated with the calculated ones are shown and we see that 
they are very well reproduced by the theory.

As regards the spectroscopic factors, the calculated values for both nuclei  evidence a strong fragmentation of the   single-particle strength with 
respect to $^{135}$Te and $^{137}$Xe.   We find, in fact, that in $^{135}$Sb 
only for three  $J^{\pi}$, which  reduce to two in $^{137}$Sb, there is one   state
with a spectroscopic factor larger than 0.4. The remaining strength in both nuclei is shared between many 
states.  We have seen in Sec. III.A that for the $N=83$ isotones this occurs 
only for $J^{\pi}=5/2^-$ in $^{137}$Xe.

\begin{table}
\caption{Calculated energies and spectroscopic factors for states in 
$^{135}$Sb.
The available experimental data are reported for comparison (see text for details).} 
\begin{ruledtabular}
\begin{tabular}{ccccc}
\multicolumn{3} {c} {Calc.}  &  \multicolumn{2} {c}
 {Expt.} \\
 \cline{1-3}    \cline{4-5}
$J^{\pi}$& E(MeV) & $C^{2}S$ & $J^{\pi}$ & E(MeV) \\
\colrule
($1/2^+$)$_{1}$ & 0.659 & 0.07 & ($1/2^{+}$) &  0.523 \\
($1/2^+$)$_{11}$ & 2.880& 0.07 &  &    \\
($1/2^+$)$_{12}$ & 3.119 & 0.32 &  &    \\
($1/2^+$)$_{15}$ & 3.476& 0.17 &  &    \\
($3/2^+$)$_{1}$ & 0.497 & 0.07 & ($3/2^{+}$)  &  0.440 \\
($3/2^+$)$_{3}$ & 1.320 & 0.07 &  &  \\
($3/2^+$)$_{12}$ & 2.600 & 0.32 &  &  \\
($3/2^+$)$_{13}$ & 2.697 & 0.10 &  &  \\
($3/2^+$)$_{14}$ & 2.768 & 0.09 &  &  \\
($5/2^+$)$_{1}$ & 0.387 & 0.42 & ($5/2^{+}$) &  0.282 \\
($5/2^+$)$_{2}$ & 0.928 & 0.23 &  &   \\
($5/2^+$)$_{5}$ & 1.657 & 0.09 &  &   \\
($5/2^+$)$_{8}$ & 1.950 & 0.14 &  &   \\
($7/2^+$)$_{1}$ & 0.000 & 0.74 & ($7/2^{+}$)  &  0.000 \\
($7/2^+$)$_{2}$ & 0.944 & 0.18 &   &   \\
($11/2^-$)$_{1}$ & 2.652 & 0.52 &  &   \\
($11/2^-$)$_{2}$ & 3.132 & 0.11 &  &   \\
($11/2^-$)$_{5}$ & 3.522 & 0.21 &  &   \\
\end{tabular}
\end{ruledtabular}
\label{135sb}
\end{table}

\begin{table}
\caption{Calculated energies and spectroscopic factors for states in $^{137}$Sb.} 
\begin{ruledtabular}
\begin{tabular}{ccc}
\multicolumn{3} {c} {Calc.}  \\
 \cline{1-3}    
$J^{\pi}$& E(MeV) & $C^{2}S$  \\
\colrule
($1/2^+$)$_{1}$ & 0.403 & 0.11  \\
($3/2^+$)$_{1}$ & 0.333 & 0.12  \\
($5/2^+$)$_{1}$ & 0.186 & 0.53  \\
($5/2^+$)$_{10}$ & 1.538 & 0.09  \\
($7/2^+$)$_{1}$ & 0.000 & 0.71 \\
($7/2^+$)$_{2}$ & 0.913 & 0.09 \\
($11/2^-$)$_{1}$ & 2.587 & 0.38     \\
($11/2^-$)$_{2}$ & 2.848 & 0.15 \\
\end{tabular}
\end{ruledtabular}
\label{137sb}
\end{table}

The spreading of the single-proton strength in the $Z=51$ isotopes can be 
better seen through the cumulative sum of the spectroscopic  factors. This
is shown in Fig.~\ref{fig2} for each single-proton orbit  as a 
function  of the number of states included in the sum.
We see that the fragmentation is particularly strong for the $s_{1/2}$ and
$d_{3/2}$ strengths, which in $^{137}$Sb do not reach  significant values even 
when including the 15 lowest-lying states. 

\begin{figure}

\centering

\includegraphics [scale=0.8]{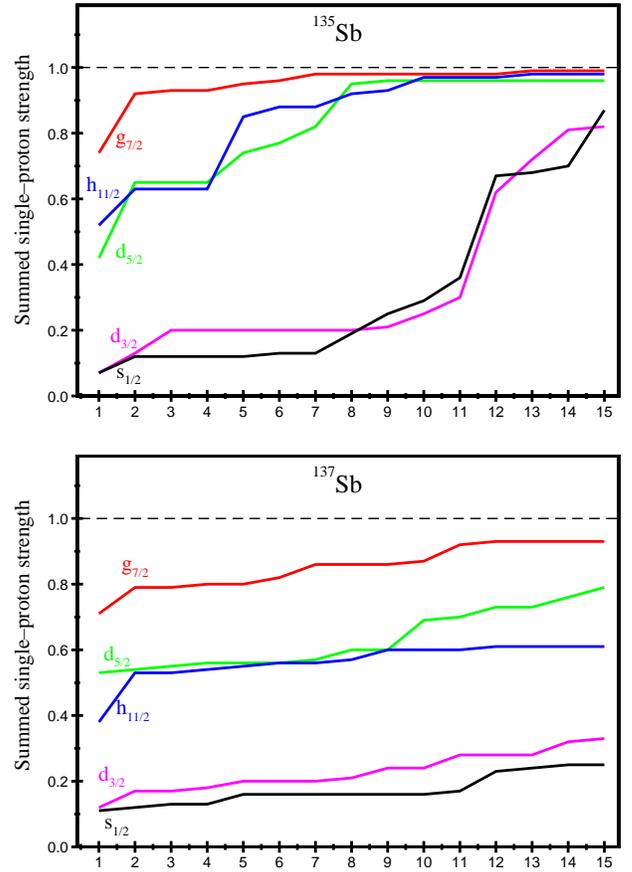}

\caption{Summed  spectroscopic factors for  $^{135}$Sb and  $^{137}$Sb as a function of 
the  included states (see text for details).}\label{fig:2}

\end{figure}

As a  direct consequence of the large fragmentation induced by addition of 
pairs of 
neutrons, the single-particle spectrum 
of $^{133}$Sb is substantially modified, as shown 
in Fig. \ref{fig3} which  is the counterpart of Fig.~\ref{fig1}.
This occurs already in $^{135}$Sb with only two more neutrons, the most noticeable change 
being the lowering of the $5/2^+$ state by  about 600 keV. 

The low position of the $5/2^+$ state in $^{135}$Sb has been indeed  in focus  of great
attention~\cite{Coraggio05,Korgul07}. 
In some studies,  this was traced   to a decrease of the proton 
$d_{5/2} - g_{7/2}$ 
spacing  produced by the two neutrons beyond the $N=82$ shell closure.
On the other hand, our predicted spectroscopic factor, 0.42, evidences that this state  has no strong single-particle nature, namely 
non-negligible components with seniority larger than one  are contained in its
wave function. This was discussed in detail in Ref.~\cite{Coraggio05}, where we 
showed that the seniority-one state 
$|\pi d_{5/2} (\nu f_{7/2})^{2}_{J=0}>$,
with an unperturbed energy of 0.962 MeV, is pushed down by the neutron-proton interaction getting admixed  with the seniority-three state
$|\pi g_{7/2}(\nu f_{7/2})^{2}_{J=2}>$, which lies at 0.368 MeV.

\begin{figure}

\centering

\includegraphics [scale=0.8]{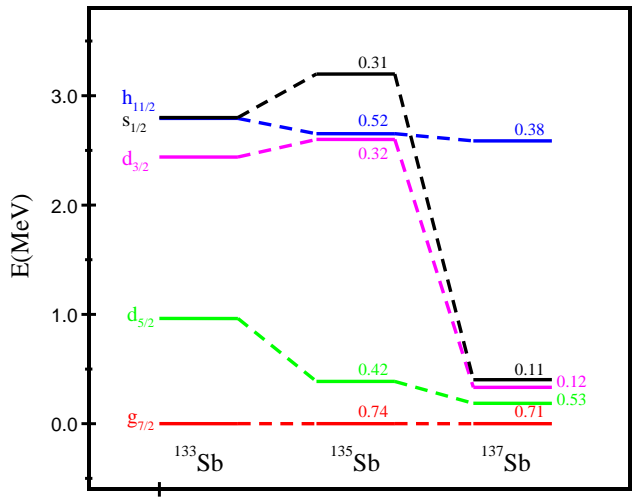}

\caption{Calculated energy levels with corresponding spectroscopic factors  in 
the $Z=51$ isotopes $^{135}$Sb and  $^{137}$Sb (see text for details). Single-proton  levels for $^{133}$Sb are also reported.} \label{fig:3}

\end{figure}

This mechanism explains indeed the fragmentation of the single-particle strength we predict for almost all states in  both $^{135}$Sb and $^{137}$Sb. By the same token, we see
 why no strong fragmentation is found for the 
$N=83$ isotones, most of  the seniority-one states lying, in this case,
sufficiently lower in energy with respect to the seniority-three states.
This difference in the evolution of the single-proton versus single-neutron 
states may be then traced to the different  pairing force acting between 
protons and neutrons.

It is indeed  a feature  of our effective interaction to have different pairing properties for protons above $Z=50$ and neutrons above $N=82$, which results 
in a  ``normal'' pairing for protons and a significantly weaker pairing for neutrons. In our derivation of the effective interaction, this arises from the 
second-order core polarization effects, the main role being played by 
one-particle-one hole excitations \cite{Covello12}. Experimentally, 
the reduction of neutron 
pairing is clearly manifested  by the large difference
between the proton and neutron  gap in $^{134}$Te and
$^{134}$Sn, the latter being about 0.5 MeV smaller than the former.

\section{Effective single-particle energies}

In this section we discuss the single-proton and -neutron energies which
can be associated to the underlying mean field when neutron and proton 
pairs are added to $^{133}$Sb or $^{133}$Sn, respectively.
These quantities are  the effective 
single-particle energies (ESPEs)~\cite{Umeya06,Duguet12}

\begin{equation}
\bar{\epsilon}_{j_{\rho}}=\epsilon_{j_{\rho}} + \sum_{j_{\rho'}} 
V^{M}(j_{\rho} j_{\rho'}) N_{j_{\rho'}},
\label{one}
\end{equation}

\noindent {where  $\rho$ and $\rho'$ stand for  neutron and proton index, respectively, or viceversa. For a given $j_{\rho}$,
$\epsilon_{j_{\rho}}$  denotes the corresponding energy in  the 
one-valence system and $N_{j_{\rho}}$ the occupation number  in the ground state of the even-even system, while}

\begin{equation}
V^{M}(j_{\rho}j_{\rho'})= \frac{\sum_{J} (2J+1)<j_{\rho}j_{\rho'};J |V|j_{\rho}j_{\rho'};J>} {\sum_{J}(2J+1)}
\end{equation}

\noindent{ is the monopole component of the neutron-proton interaction.}

It is worth noting  that the  ESPEs coincide \cite{Umeya06} with the energy 
centroids  defined  in ~\cite{Baranger70}
more than 40 years ago. The 
practical use of the latter  definition, however, requires  knowledge of 
all energies and spectroscopic factors  corresponding to a given angular 
momentum, whereas  expression~(\ref{one}) makes the computation of the ESPEs 
straightforward.

\begin{figure*}

\centering

\includegraphics [scale=0.8]{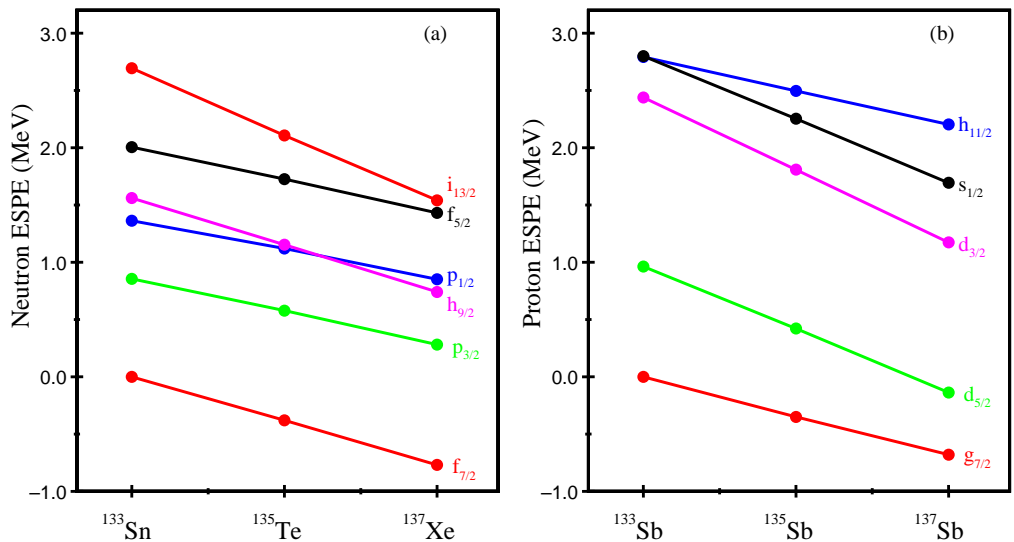}

\caption{Evolution of (a) neutron and (b) proton ESPEs.} \label{fig:4}

\end{figure*}

We have employed Eq.~(\ref{one}) to calculate the  neutron  and proton ESPEs whose
values  are shown in  Fig. \ref{fig4}. By examining 
the various terms of~(\ref{one}), 
we see that only some of them give an important contribution to these
energies. In particular, for each effective single-neutron energy 
in the $N=83$ isotones, the dominant contribution is that corresponding  
to $\pi g_{7/2}$   due to its large 
occupancy. We predict, in fact, that 1.6 out of  2 
protons are in the $\pi g_{7/2}$ orbit for $^{134}$Te, and 3.6 out of  4 
for $^{136}$Xe. Note that our predicted occupancies for the ground state of
$^{136}$Xe are in good agreement with the results obtained from the experiment of Ref.~\cite{Wildenthal71}.
Similar results are found   for the Sb isotopes as regards the occupancy of the
neutron $f_{7/2}$ orbit.  The monopole matrix elements entering  the dominant 
terms for the neutron and proton ESPEs are therefore   
$V^{M}(j_{\nu} \pi g_{7/2})$ and $V^{M}(\nu f_{7/2}  j_{\pi})$, respectively, which are reported in Tables \ref{ESPEn} and \ref{ESPEp}. 

\begin{table}
\caption{Neutron-proton monopole components $V^{M}(j_{\nu} \pi g_{7/2})$ (in MeV).} 
\begin{ruledtabular}
\begin{tabular}{cccccc}   
$\nu p_{1/2}$ & $\nu p_{3/2}$& $\nu f_{5/2}$& $\nu f_{7/2}$ & $\nu h_{9/2}$ &$\nu i_{13/2}$ \\
\colrule
-0.09& -0.11 & -0.11&-0.18&-0.20&-0.33  \\
\end{tabular}
\end{ruledtabular}
\label{ESPEn}
\end{table}

\begin{table}
\caption{Neutron-proton monopole components $V^{M}(\nu f_{7/2} j_{\pi})$ (in MeV).} 
\begin{ruledtabular}
\begin{tabular}{cccccc}   
$\pi s_{1/2}$ & $\pi d_{3/2}$& $\pi d_{5/2}$& $\pi g_{7/2}$ & $\pi h_{11/2}$  \\
\colrule
-0.27& -0.35 & -0.28&-0.18&-0.13  \\
\end{tabular}
\end{ruledtabular}
\label{ESPEp}
\end{table}

From Fig. \ref{fig4}, we see that all the neutron and proton ESPEs  go down  
when pairs of nucleons are added. However, the decrease in energy  is not the 
same for the various levels leading to single-particle spectra substantially 
different from those of the one valence-particle nuclei. For instance, the five neutron ESPEs above the lowest one tend to lie in a smaller energy interval, the $i_{13/2}-f_{5/2}$ spacing reducing from the initial value of 700  to 100 keV in $^{137}$Xe. On the other hand,  the spacing between the lowest and the highest proton levels remains  practically unchanged while  the three other levels  move toward the $g_{7/2}$ with an overall downshift of about 500 keV in $^{137}$Sb. 

In this context, it should be mentioned that the evolution of shell structure has received great attention in recent years owing to new available data on exotic nuclei, which have evidenced the appearance or disappearance of magic numbers. Several theoretical studies have then been carried out to try to  identify the  mechanism behind the variation of single-particle energies. In particular, the role of the monopole components of the spin-isospin and tensor interaction has been pointed out, the latter producing an attractive/repulsive
effect for two unlike nucleons with  opposite/aligned spin orientation ~\cite{Otsuka01,Otsuka05}.
By using the spin-tensor decomposition of the interaction, the interplay between the tensor force and 
other components of the neutron-proton interaction
has been investigated~\cite{Smirnova12}. This has provided evidence for the importance  of both the central 
and tensor force  in determining the evolution of ESPEs, the latter playing a dominant role for 
the energy difference of spin-orbit partners.

It would be clearly desirable to perform  such a decomposition also in the present study to 
clarify how the behavior of the ESPEs shown in Fig. \ref{fig4} is related to the various 
terms of our effective interaction.  This is, unfortunately, not feasible since our  
shell-model space, on which the interaction is defined,  does not  contain for each orbital 
angular momentum the two spin-orbit partners. This is, in fact, a prerequisite to perform the spin-tensor decomposition.
We cannot therefore disentangle the effects of the central and tensor force, 
so as to
unambiguously assess the role of the various terms of the effective neutron-proton interaction in determining the evolution of the ESPEs. 

From Tables \ref{ESPEn} and \ref{ESPEp}, however, we see  that for each pair of neutron/proton 
spin-orbit partners the monopole matrix element corresponding to the orbit with opposite spin 
orientation relative to  proton $g_{7/2}$/neutron   $f_{7/2}$ is the most attractive.  
This accounts for the variation  in the $f_{5/2}-f_{7/2}$ and $p_{1/2}-p_{3/2}$ neutron spacings, 
as well as  in the $d_{3/2}-d_{5/2}$ proton one, and may be traced to the effect of the tensor component of our effective interaction \cite{Otsuka05,Smirnova12}. We also predict a significant decrease in
the neutron $i_{13/2}-h_{9/2}$ spacing, which
comes  essentially from  $V^{M}(\nu i_{13/2} \pi g_{7/2})$ being more attractive  than 
$V^{M}(\nu h_{9/2} \pi g_{7/2})$.  The less attraction of the latter  may be again
attributed to the tensor force, the neutron $h_{9/2}$ orbit, unlike the  $i_{13/2}$ one, being spin-down as  the proton $g_{7/2}$ orbit \cite{Otsuka05}. In this connection, it is worth mentioning that this force 
has been  shown to be responsible of the reduction in 
the separation of these two orbits  in the $N=83$ isotones \cite {Kay11,Otsuka05}.

\section{Concluding remarks}

In this work, we have performed shell-model calculations for the four neutron-rich nuclei
$^{135}$Sb, $^{137}$Sb, $^{135}$Te and $^{137}$Xe. The main aim of our study was 
to follow the evolution of the single-proton and -neutron states outside doubly magic 
$^{132}$Sn when adding neutron and proton pairs, respectively. 
It is worth emphasizing that in our calculations we have consistently employed the same Hamiltonian which has given a very good description of other nuclei beyond $^{132}$Sn
\cite{Coraggio09,Covello09,Covello11} without use of any adjustable parameters.

We have calculated energies, spectroscopic factors and effective 
single-particle energies for states of all the four nuclei considered and 
compared our results with the available experimental data which are, however, 
rather scanty. In particular, no spectroscopic factors are available for $^{135}$Te and
$^{135}$Sb while no spectroscopic information at all exists for $^{137}$Sb.  This amounts to say that this study is largely predictive in nature and is meant to stimulate, and be of guide to, experimental efforts with RIBs in the beyond $^{132}$Sn region. 

Of noteworthy interest is the stronger fragmentation of the single-particle strengths predicted for $^{135}$Sb and 
$^{137}$Sb as compared to that for $^{135}$Te and $^{137}$Xe. This stems 
essentially from the fact that the neutron pairing beyond $N=82$ is 
significantly smaller than the proton pairing beyond $Z=50$. We have shown in 
\cite{Covello12} that this  difference in the proton and neutron effective interaction finds its origin in core-polarization effects.

As shown in Tables II, III and IV, our results are in very good agreement with the measured excitation energies in $^{135}$Te, $^{135}$Sb and $^{137}$Xe as well as with the spectroscopic factors extracted from the data for the latter. This makes us confident in the predictive power of our calculations. 

\begin{acknowledgments}
This work has been supported in part by the Italian Ministero  dell'Istruzione e della Ricerca  (MIUR) under PRIN 2009.
\end{acknowledgments}

\end{document}